\documentclass{webofc}
\usepackage[varg]{txfonts}   
\usepackage{graphicx}
\usepackage{hyperref}
\usepackage{lineno}
\usepackage{color}
\usepackage{caption}
\usepackage{multirow}
\usepackage{graphics}
\usepackage{epstopdf}
\usepackage{amssymb}
\usepackage{xspace}
\usepackage{amsmath}
\usepackage{upgreek}
\usepackage{hyperref}
\newcommand{\ptt}{\ensuremath{p_{\mathrm{T}}^{\rm trig}}\xspace}
\newcommand{\pl}{\ensuremath{p_{\mathrm{T}}^{\rm leading}}\xspace}
\newcommand{\pt}{\ensuremath{p_{\rm T}}\xspace}
\newcommand{\pp}{pp\xspace}
\newcommand{\pPb}{\mbox{p--Pb}\xspace}
\begin{document}
%
\title{Particle production as a function of underlying-event activity and very forward energy with ALICE}
\author{\firstname{Feng} \lastname{Fan} \thanks{\email{feng.fan@cern.ch}} (for the ALICE Collaboration)}

\institute{Central China Normal University, Wuhan, Hubei, 430079, China}

\abstract{
Measurements of charged-particle production in \pp and \pPb collisions at $\sqrt{s}_{\rm NN}=5.02$\,TeV in the toward, away and transverse regions are discussed. These three regions are defined event by event relative to the track with the largest transverse momentum (\ptt). The transverse region is sensitive to the underlying event (UE), but it also includes contributions from initial- and final-state radiation (ISR-FSR). Therefore, it is further subdivided in two regions, defined according to their relative multiplicity: trans-max (the sub-transverse region with the largest multiplicity) and trans-min (the sub-transverse region with the smallest multiplicity) regions which are sensitive to UE and ISR-FSR, respectively. KNO-like scaling properties are explored in the three defined transverse regions. Finally, the relationship between \ptt and the energy detected in a region close to beam rapidity (very forward energy) is reported.
}
\maketitle

\section{Introduction}
\label{intro}
In models incorporating multi-parton interactions (MPI), particles produced in the hard scattering (jet) are accompanied by particles from additional parton-parton interactions, as well as from the proton break-up~\cite{Sjostrand:1987su}. This component of the collision makes up the underlying event (UE). The traditional UE analysis focuses on the study of particles in three topological regions depending on their azimuthal angle relative to the leading particle ($|\Delta\varphi| = |\varphi - \varphi^{\rm trig}|$), which is the one with the highest transverse momentum in the event (\ptt or \pl) at mid-pseudorapidity ($|\eta|<0.8$). The toward region ($|\Delta\varphi| < \pi/3$\,rad) contains the primary jet and UE, while the away region ($\pi/3$\,rad < $|\Delta\varphi| < 2\pi/3$\,rad) contains the fragments of the recoil jet and UE. In contrast, the transverse region ($|\Delta\varphi| > 2\pi/3$\,rad) is dominated by the UE dynamics, but it also includes contributions from initial- and final-state radiation (ISR-FSR)~\cite{Bencedi:2021tst,Ortiz:2021gcr}. A study based on MC simulations shows that at the LHC energies the multiplicity distributions in the transverse region obey a Koba-Nielsen-Olesen (KNO) scaling~\cite{Ortiz:2017jaz}. In this contribution, the transverse region is further subdivided into trans-max and trans-min, defined accoriding to their relative multiplicity, in order to increase the sensitivity to ISR-FSR and UE effects, respectively. Therefore, measurements of multiplicity distributions in the trans-min region at different centre-of-mass energy allows for the study of the role of MPI to produce such scaling properties. Another way to characterise the event is to look into a region in the very forward pseudorapidity ($|\eta|>8.8$) using the ALICE zero degree calorimeters (ZDC), which measure the beam remnants. Moreover, the relationship between the very forward energy and \ptt is complementary to the UE measurements, and provides direct insights into the initial stages and the subsequent evolution of the collision. In this contribution, a set of new measurements aimed at understanding the role of MPI in hadronic interactions at the LHC is presented~\cite{ALICE:2022fnb,ALICE:2022qxg}. 
\begin{figure}[h]
\centering
\includegraphics[width=15pc]{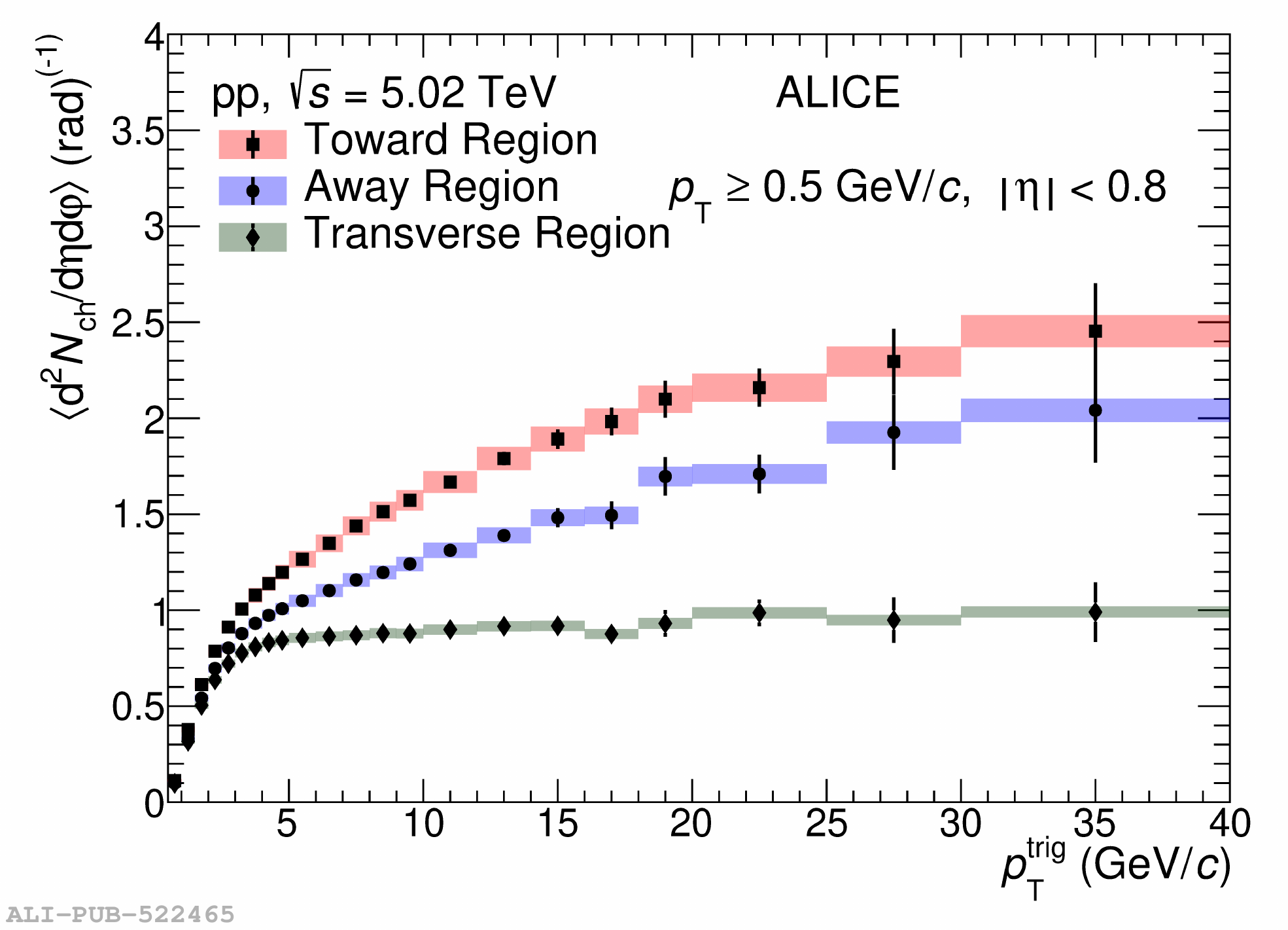}
\includegraphics[width=15pc]{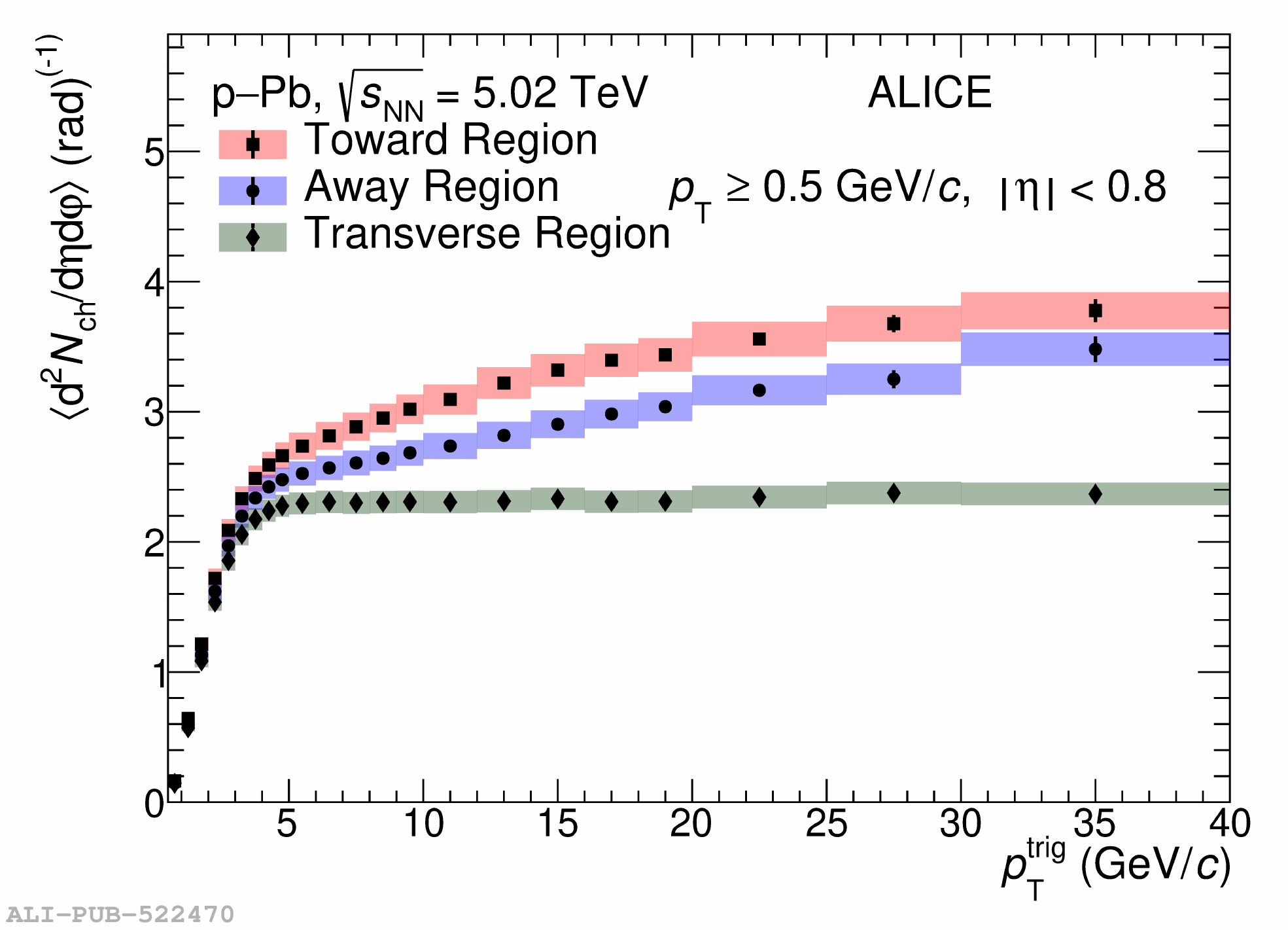}
\caption{Number density as a function of \ptt measured in \pp (left) and \pPb (right) collisions at $\sqrt{s_{\rm NN}}=5.02$\,TeV. Results for the toward, transverse, and away regions are displayed. The boxes and the error bars represent the systematic and statistical uncertainties, respectively.}
\label{fig:1}  
\end{figure}

\section{Results and discussion}
Figure~\ref{fig:1} shows the number density as a function of \ptt measured in \pp and \pPb collisions at $\sqrt{s_{\rm NN}}=5.02$\,TeV. Results are presented for the toward, transverse and away regions. For both collision systems, the distributions behave in the same way. At low \ptt ($<5$\,GeV/$c$), the number densities rapidly increase with \ptt for those three topological regions, while at higher \ptt ($>5$\,GeV/$c$), they still increase but less steep in the toward and away regions, whereas in the transverse region they tend to saturate. This saturation is expected in models  including an impact parameter dependence such that the requirement of the presence of a high-\pt particle in a \pp collision biases the selection of collisions towards those with a smaller impact parameter~\cite{Strikman:2011cx}. Moreover, the increase observed in the toward and away regions at high \pt is due to the contributions from jets as well as ISR and FSR. The UE contributions to the toward and away regions are larger in \pPb than in \pp collisions.
\begin{figure}[h]
\centering
\includegraphics[width=14pc]{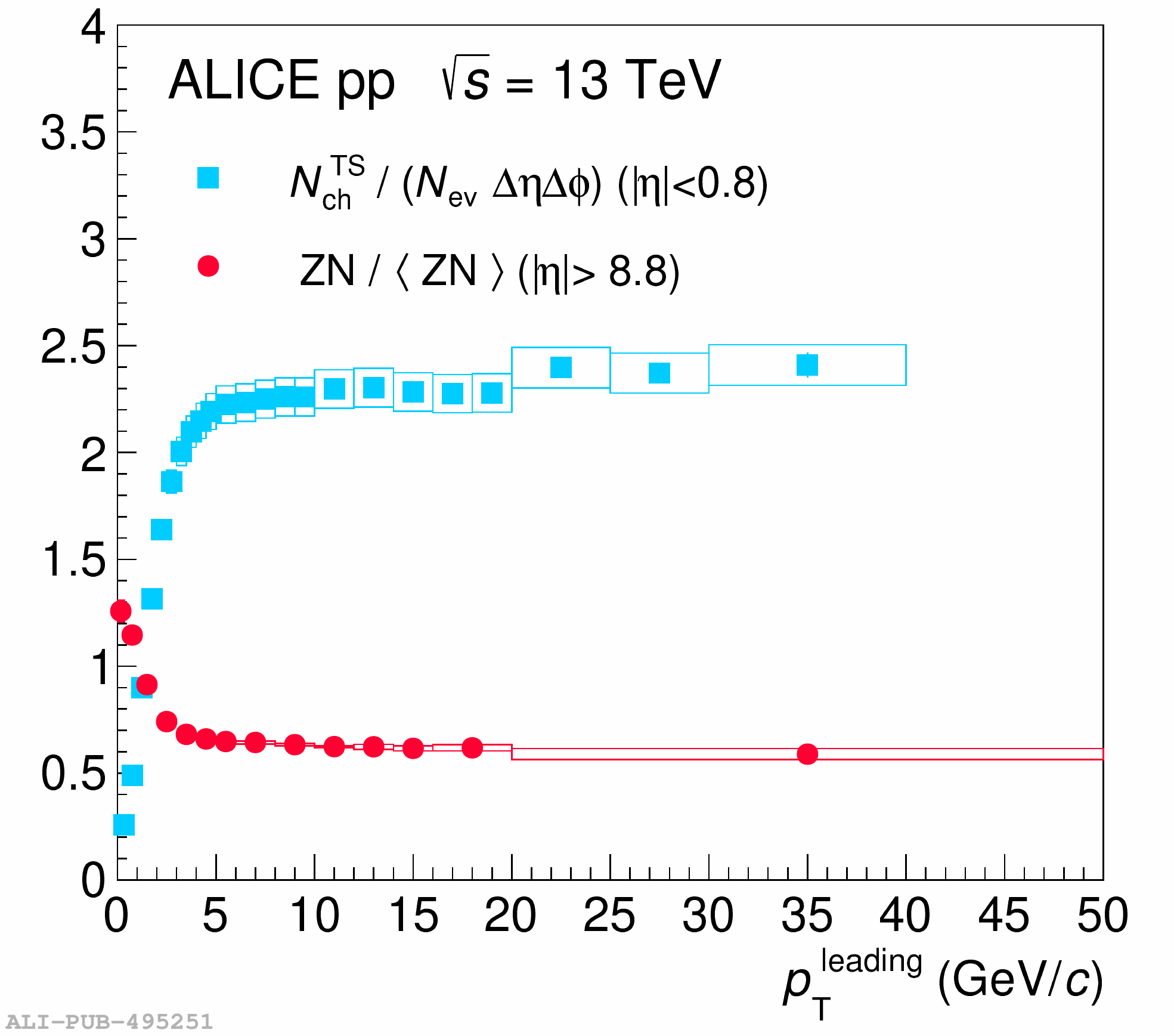}
\caption{Number density (azure squares) in the transverse region~\cite{ALICE:2019mmy} and self-normalised ZN signal (red circles) as a function of \pl at mid-pseudorapidity in \pp collisions at $\sqrt{s}=13$\,TeV.}
\label{fig:2}     
\end{figure}

Figure~\ref{fig:2} shows the comparison between the number density in the transverse region~\cite{ALICE:2019mmy} and the self-normalised neutral ZDC energy (ZN) as a function of \pl at mid-pseudorapidity in \pp collisions at $\sqrt{s}=13$\,TeV. The very forward energy reaches its minimum value at \pl$\approx5$\,GeV/$c$ and then remains constant for increasing \pl. The saturation occurs at the same \pt value observed for UE quantities. The result corroborates the interpretation in term of a bias towards collisions with small impact parameter, because the correlation between central and forward pseudorapidity can only be attributed to the initial stage of the collisions. 
\begin{figure}[h]
\centering
\includegraphics[width=15pc]{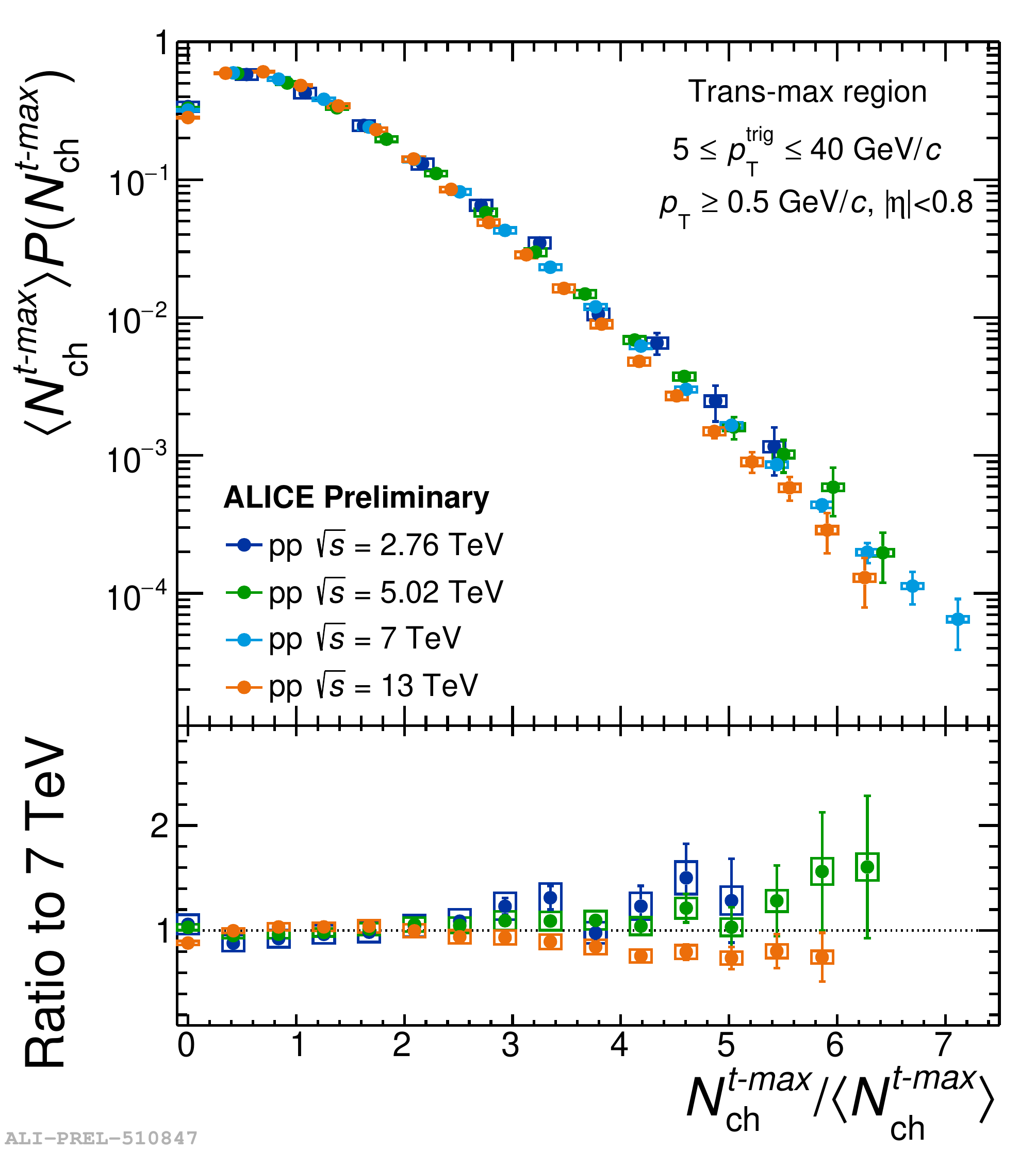}
\includegraphics[width=14.8pc]{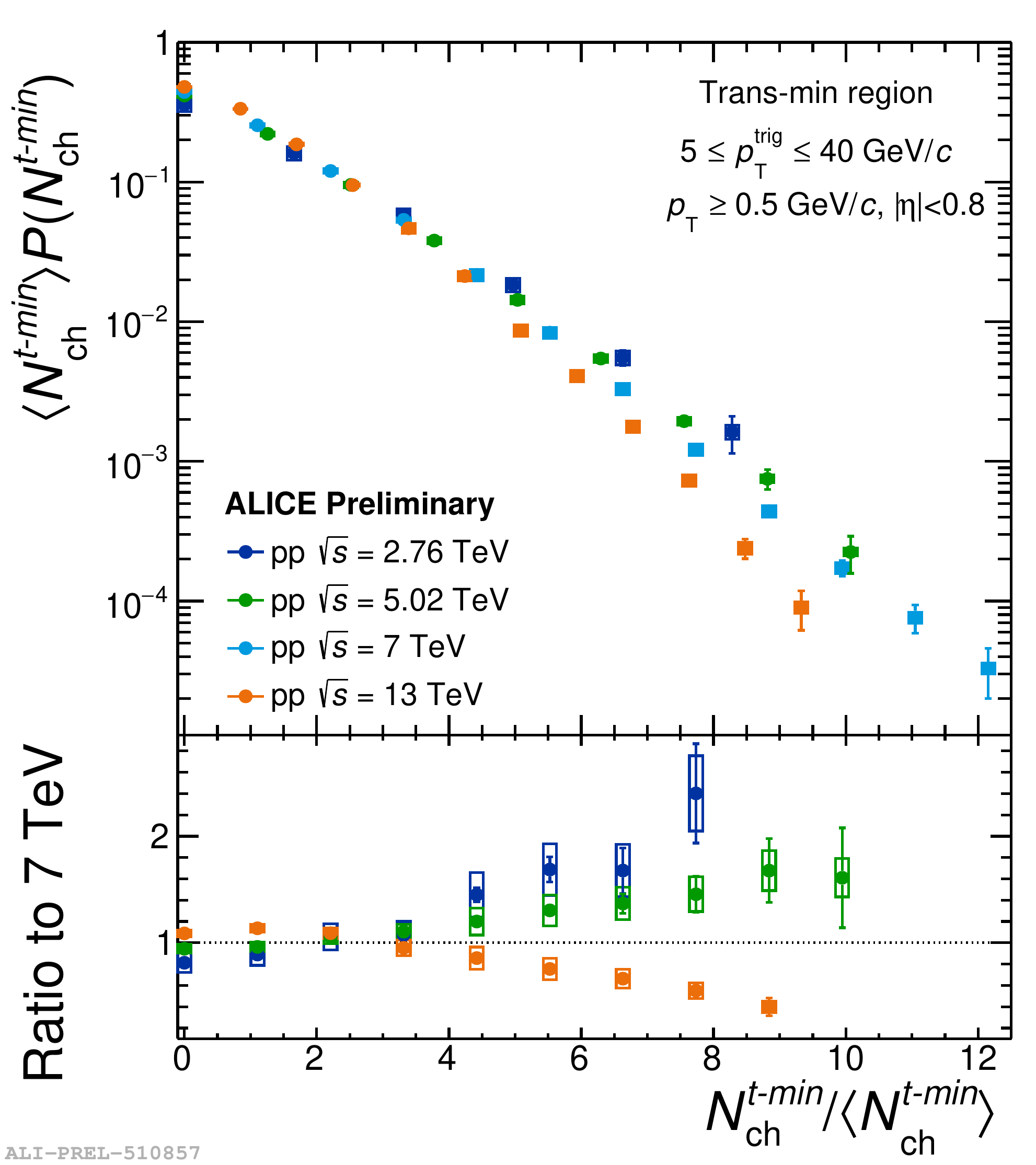}
\caption{Top: multiplicity distributions in KNO variables in the tran-max (left) and trans-min (right) regions for \pp collisions at $\sqrt{s}=2.76$, 5.02, 7 and 13\,TeV. Bottom: the KNO multiplicity distributions are normalized to that for \pp collisions at $\sqrt{s} = 7$\,TeV. The boxes and the error bars represent the systematic and statistical uncertainties, respectively.}
\label{fig:3}  
\end{figure}

In a Monte Carlo study of UE~\cite{Ortiz:2017jaz}, it was shown that the multiplicity distributions in the transverse region ($|\eta|<2.5, \pt>0$\,GeV/$c$) at the plateau obey a KNO scaling at the LHC energies. The KNO scaling was expected in a model which assumes that a single \pp collision results from the superposition of a given number of elementary partonic collisions emitting independently~\cite{DiasdeDeus:1997ui}. Therefore, MPI could produce such an effect. However, the transverse region not only includes contributions from UE but also from ISR-FSR. In order to investigate the KNO-like scaling properties, a further treatment of the transverse side is implemented. The transverse region can be further divided in two regions:
\begin{itemize}
\item transverse-I: $\pi/3<\Delta\varphi<2\pi/3$
\item transverse-II: $\pi/3<-\Delta\varphi<2\pi/3$
\end{itemize}
These two distinct regions are characterized in terms of their relative charged-particle multiplicities. Trans-max and trans-min regions refer to the sub-transverse region (I or II) with the largest and smallest charged-particle multiplicity which have an enhanced sensitivity to ISR-FSR and UE, respectively~\cite{Bencedi:2021tst, CDF:2015txs}. 
 
Figure~\ref{fig:3} shows the charged-particle multiplicity distributions in KNO variables for \pp collisions at $\sqrt{s}=2.76$, 5.02, 7 and 13\,TeV. Results are presented for the trans-max and trans-min regions. In the trans-max region, within 20\%, the KNO-like scaling is observed in a wider range of multiplicity ($0<z<4$) relative to the results reported in Ref.~\cite{Ortiz:2021gcr}, while for higher $z$ values ($z>4$) the scaling is broken. It is worth noticing that for trans-max both contributions are considered: UE and ISR-FSR. If the effect of ISR-FSR is suppressed, i.e., exploiting the features of trans-min region, the KNO-like scaling also holds for $0<z<4$, and then for $z>4$ the KNO-like scaling is still broken but a higher $z$ reach is achieved, especially for $z>6$, a lager violation is observed. Events with high-multiplicity jets can contribute to the large violation of the scaling properties. It was observed that for $z>3$, the number of uncorrelated seeds (or MPI) deviate from the linear trend suggesting the presence of high-multiplicity jets~\cite{ALICE:2013tla, Ortiz:2021peu}. 

\section{Conclusions}
The number density as a function of \ptt for both \pp and \pPb collisions at $\sqrt{s_{\rm NN}}=5.02$\,TeV shows a remarkably similar behavior, such as in the transverse region the distributions increase with \ptt until about 5\,GeV/$c$ where it saturates. On the other hand, the very forward energy as a function of \pl in \pp collisions at $\sqrt{s}=13$\,TeV decreases with increasing \pl and then saturates at the same \pl ($\approx5$\,GeV/$c$). These two saturation effects are  commonly interpreted as a bias towards collisions with small impact parameter. In addition, in the trans-max and trans-min regions the charged-particle multiplicity distributions exhibit a KNO-like scaling for $0<z<4$, which is expected for a single pp collision involving a large number multi-partonic interactions emitting independently. And for $z>4$ the scaling is broken. For the MPI-sensitive region, a higher $z$ reach is achieved, especially for $z>6$, a larger violation is observed, which may be attributed to high-multiplicity mini jets.

\section{Acknowledgement}
This work has been supported by CONACyT under the Grants CB No. A1-S-22917 and CF No. 2042.

\bibliography{main}

\begin{thebibliography}{12}

\bibitem{Sjostrand:1987su}
T.~Sj{\"o}strand, M.~van Zijl, Phys. Rev. D \textbf{36}, 2019 (1987)

\bibitem{Bencedi:2021tst}
G.~Bencedi, A.~Ortiz, A.~Paz, Phys. Rev. D \textbf{104}, 016017 (2021),
  \texttt[arXiv]{2105.04838}

\bibitem{Ortiz:2021gcr}
A.~Ortiz, Phys. Rev. D \textbf{104}, 076019 (2021), \texttt[arXiv]{2108.08360}

\bibitem{Ortiz:2017jaz}
A.~Ortiz, L.~Valencia~Palomo, Phys. Rev. D \textbf{96}, 114019 (2017),
  \texttt[arXiv]{1710.04741}

\bibitem{ALICE:2022fnb}
S.~Acharya et~al. (ALICE) (2022), \texttt[arXiv]{2204.10389}

\bibitem{ALICE:2022qxg}
S.~Acharya et~al. (ALICE) (2022), \texttt[arXiv]{2204.10157}

\bibitem{Strikman:2011cx}
M.~Strikman, Acta Phys. Polon. B \textbf{42}, 2607 (2011),
  \texttt[arXiv]{1112.3834}

\bibitem{ALICE:2019mmy}
S.~Acharya et~al. (ALICE), JHEP \textbf{04}, 192 (2020),
  \texttt[arXiv]{1910.14400}

\bibitem{DiasdeDeus:1997ui}
J.~Dias~de Deus, C.~Pajares, C.A. Salgado, Phys. Lett. B \textbf{408}, 417
  (1997), \texttt{hep-ph/9705425}

\bibitem{CDF:2015txs}
T.A. Aaltonen et~al. (CDF), Phys. Rev. D \textbf{92}, 092009 (2015),
  \texttt[arXiv]{1508.05340}

\bibitem{ALICE:2013tla}
B.~Abelev et~al. (ALICE), JHEP \textbf{09}, 049 (2013),
  \texttt[arXiv]{1307.1249}

\bibitem{Ortiz:2021peu}
A.~Ortiz, E.A. Zepeda, J. Phys. G \textbf{48}, 085014 (2021),
  \texttt[arXiv]{2101.10274}

\end{thebibliography}
\end{document}